\documentclass[11pt]{article} \usepackage{latexsym} \usepackage{graphicx} \usepackage{enumerate} \usepackage{delarray}
\usepackage{tabularx} \usepackage{bm} \usepackage{amssymb}
\usepackage[small,labelsep=period,figurename=Fig. ]{caption}
\begin{document}

\title{{\LARGE Network versus portfolio structure in financial systems}} \author{{\Large {Teruyoshi Kobayashi}\thanks{
Graduate School of Economics, Kobe University, 2-1 Rokkodai-cho, Nada-ku, Kobe 657-8501, Japan. E-mail:
kobayashi@econ.kobe-u.ac.jp. Tel, fax: +81-78-803-6692.}}} \date{{\large {July, 2013}}} \maketitle
\begin{abstract}
The question of how to stabilize financial systems has attracted considerable attention since the global financial crisis of 2007-2009. 
Recently, Beale et al. (``Individual versus systemic risk and the regulator\rq{}s dilemma\rq\rq{}, \textit{Proc Natl Acad Sci USA} 108: 12647-12652, 2011) demonstrated that higher portfolio diversity among banks would reduce systemic risk by decreasing the risk of simultaneous defaults at the expense of a higher likelihood of individual defaults. In practice, however, a bank default has an externality in that it undermines other banks\rq{} balance sheets. This paper explores how each of these different sources of risk, simultaneity risk and externality, contributes to systemic risk. The results show that the allocation of external assets that minimizes systemic risk varies with the topology of the financial network as long as asset returns have negative correlations. In the model, a well-known centrality measure, PageRank, reflects  an appropriately defined ``infectiveness\rq\rq{} of a bank. An important result is that the most infective bank need not always be the safest bank. Under certain circumstances, the most infective node should act as a firewall to prevent large-scale collective defaults.  The introduction of a counteractive portfolio structure will significantly reduce systemic risk. 
\end{abstract} \thispagestyle{empty} \newpage \pagenumbering{arabic}

\section{Introduction}
 In the past 5 years, essential differences between micro- and macro-prudential policies have been emphasized by many academic researchers, central bankers and regulators (Beale et al., \cite{beale11}, Upper, \cite{upper11}, May and Arinaminpathy, \cite{may10}).
Since the failures of Bear Stearns and Lehman Brothers in 2008, there seems to have been wide agreement that micro-prudential policies do not necessarily add up to a macro-prudential policy. An ongoing challenge for regulators is to find the best macro-prudential policy that stabilizes the whole financial system, minimizing the expense of individual bank failures. 

One of the main objectives of  recent studies on systemic risk has been to gain a theoretical understanding of how the topology of the financial network affects the likelihood of a financial crisis.  By employing basic concepts developed in network theory, such as percolation, assortativity and clustering, those studies attempt to extract the necessary conditions for a stable financial system by conducting simulations under various patterns of network topology (Lenzu and Tedeschi, \cite{lenzu12}, Gai and Kapadia, \cite{gai10}, Gai et al., \cite{gai11},
Nier et al., \cite{nier07}, Kyriakopoulos et al., \cite{kyriakopoulos09}).  The basic idea behind these studies is that a financial network must be robust enough to withstand default cascades. That is, the extent of an externality to which a bank failure would give must be minimized.

While those studies focus mainly on the topological aspect of financial networks, the portfolio structure of individual banks is another key to understanding systemic stability.\footnote{In this paper, any type of financial institution is called a ``bank\rq{}\rq{} for brevity.}  Beale et al. \cite{beale11} recently pointed out that higher portfolio diversity among banks would reduce systemic risk by decreasing the likelihood of simultaneous defaults, which I call ``simultaneity risk\rq\rq{} throughout this paper. This puts into effect the ``regulator\rq{}s dilemma\rq\rq{} or the ``diversity-diversification trade-off\rq\rq{}, because minimizing the risk of individual failure (i.e., portfolio diversification) does not add up to minimizing systemic risk. Ibragimov et al. \cite{ibragimov11} and Wagner \cite{wagner10,wagner11} made similar arguments.

 The main purpose of this paper is to show how the systemically optimal allocation of external risk assets would vary with network structure. To do so, both the externality and the simultaneity risk need to be taken into account. In a financial market with simultaneity risk only, where all banks are completely isolated from each other, each bank is necessarily anonymous. From a macro-prudential viewpoint, it does not matter which banks should diversify their portfolios. The only concern  is how many banks should do so. However, once a network structure is introduced, the interconnected banks are not generally anonymous because the possibility of generating default cascades could differ across banks. The systemic importance of each bank can vary in accordance with its topological location in the network.\footnote{For classical studies on the information content of network, see Rashevsky \cite{rashevsky55} and Trucco \cite{trucco56}.} The problem of how many banks should be diversified is replaced with the problem of which banks should be diversified.

In the model of this paper, an appropriately defined ``infectiveness\rq\rq{} of an individual bank is closely correlated with a well-known centrality measure, PageRank. Other centrality measures, such as the degree centrality, the eigenvector centrality, and the betweenness centrality, do not provide correct information about the infectivity of a bank. It is also shown that the topological fragility of a whole network can be well captured either by network entropy or by the Herfindahl-Hirschman Index (HHI). On average, a more concentrated financial network is topologically more fragile than a decentralized network.   

An important result of the analysis is that the most infective bank need not always be the safest bank. This might be counterintuitive because, by definition, the extent of externality is the strongest for the most infective, and at the same time the biggest, bank. A basic reason for this result is that a single default of the biggest bank could be less contributive to systemic risk than a simultaneous failure of multiple banks. If collective defaults are more costly than a single default of the biggest bank, then the biggest bank should act as a firewall to prevent large-scale default cascades. More specifically, the biggest bank should hold external assets that are negatively correlated with the assets held by other banks. It is shown that the introduction of such a counteractive portfolio structure will significantly reduce systemic risk.

This paper also reveals that any topological characteristics of the financial network are not necessarily a correct measure of systemic risk. 
If each bank has independent external assets, then there is a negative (positive) relationship between network entropy (HHI) and systemic risk. However, this correlation disappears once the asset allocation is optimized. Assigning appropriate assets to appropriate banks will eliminate the differences in systemic risk stemming from the topological dissimilarity and thereby reduce the level of systemic risk.

\section{Simultaneity risk}

Systemic risk in the financial market consists of at least two independent sorts of risks: simultaneity risk and externality. 
Simultaneous defaults could occur regardless of whether or not banks have common assets or borrow from each other.
In this sense, simultaneity risk always exists if there are multiple banks in the financial market.
If banks are interconnected through an interbank market, then there is a possibility of default cascades, which amplify the impact of the initial bank failure. It can be said that a failure of an interconnected bank has an externality in that it necessarily increases the other banks\rq{} default probability. 

In this section, I focus on simultaneity risk by considering a situation in which each bank is fully isolated and there is no interbank
trading.  Beale et al. \cite{beale11} show that portfolio diversity among banks can be desirable for the financial system as a whole because it reduces the likelihood of simultaneous defaults.  I revisit their result in the following.  

Following Beale et al. \cite{beale11}, I assume that social costs, $C$, depend on the number of bank defaults, $n$: $C(n) = n^s, s \in [1, \infty ) $. The
expected cost is thus given as
$$ E[C] = \sum_{n=1}^N q(n) n^s,$$ 
where $N$ is the number of banks and $q(\cdot )$ is the probability function for the number of default banks.
The case of $s > 1$ means that simultaneous defaults of multiple banks will be more costly than a sequence of single bank failures. A rationale for  the convexity of the cost function is that the cost in the real economic sector will increase in superlinearly as the number of bank defaults increases. 
For example, the extent of liquidity hoarding would become greater as the number of bank defaults increases, making liquidity shortages for firms and households more severe.

For simplicity, suppose that there are only two kinds of external assets: bank-specific assets and fully diversified assets.  In the case of full diversity, each bank has
a bank-specific asset that is independent of any other bank-specific assets. On the other hand, full diversification requires every bank to hold
 a common fully diversified asset.    
 As a benchmark case, it is instructive to consider a situation in which each asset return follows an $i.i.d.$ Student t-distribution with degree of freedom $v$.\footnote{Kwapie$\acute{\mathrm n}$ and Dro\.{z}d\.{z} \cite{kwapien12} showed that the distributions of the stock returns for the American companies can be approximated by $q$-Gaussian distributions. Note that a t-distribution can be represented as a $q$-Gaussian distribution if the Tsallis parameter is given as $q=(v+3)/(v+1)$. A t-distribution and a $q$-Gaussian distribution converge to a normal distribution as $v \to \infty $ and $q \to 1$, respectively.}  
At one extreme, there is no benefit of portfolio diversification if $v = 1$. Diversity is thus always desirable in this case.
For $v > 1$, we need to compare the cost and the benefit of diversity with those under diversification. Let $K$ denote the variety of external assets. 
 If we restrict our attention to the case of $N=K$, in which case full diversity is made possible by the minimum number of assets, then the expected costs under full diversity are given as
\begin{equation} 
 E[C]_{diversity} = \sum_{n=0}^N {}_N\mathrm{C}_np^n(1-p)^{N-n}n^s,
\end{equation}
where the threshold of default is determined such that default occurs with probability $p$ if the bank has a single asset.  Notice that the variance of asset returns does not matter in the case of full diversity because each bank has only one asset. Since eq.(1) is equivalent to the $s$-th moment of binomial distribution 
$\mathrm{B}(N,p)$, this converges to the $s$-th moment of normal distribution $\mathrm{N}(Np,Np(1-p))$ as $N$ goes to infinity. 

At the other extreme of $v = \infty$, or if asset returns follow a normal distribution, diversification is more likely to be a good option.
If asset returns obey $i.i.d.$ normal,  $\mathrm{N}(0, \sigma^2 )$, the expected costs under full diversification are given by
\begin{eqnarray} 
 E[C]_{diversification_\infty} &=& \sum_{n=0}^N H_N^\infty (n) n^s \nonumber \\
                                             &=& H_N^\infty (N)N^s \nonumber \\
                                             &=& F_N(x) N^s ,
\end{eqnarray}
where $H_N^\infty (\cdot )$ is the probability function for the number of defaults under full diversification. The second equality follows from the fact that all banks share an identical portfolio. $F_N(\cdot )$ is the CDF of $\mathrm{N}(0, \sigma^2/N)$, and $x$ is the threshold value of the diversified asset below which all the banks go bankrupt.    
It follows that 
\begin{equation}
F_N(x)N^s = N^s\cdot\frac{\sqrt{N}}{\sigma\sqrt{2\pi}}\int_{-\infty} ^x \exp \left(  -\frac{y^2}{2\sigma^2/N}\right) dy.
\end{equation}
No matter how big $s$ is, the presence of $N$ in the exponential term suggests that the expected costs will converge to zero as $N$ goes to infinity.   
However, when $N$ is sufficiently small, an increase in $N$ will lead to higher expected costs.


 Figure 1 illustrates how the expected costs change with the size of $N (=K)$. As noted above, the benefit of diversification increases as 
the degree of freedom, $v$, goes up. According to the figure, if $v = 3$, then diversity is more desirable than diversification as long as $N < 30$.   
If $v = \infty$, on the other hand, then full diversification will be the optimal choice for $N > 4$. This reconfirms the point noted by Ibragimov et al. [9] that a fatter-tailed distribution makes diversification more dangerous and thus diversity more attractive.  The extent of simultaneity risk depends heavily on the thickness of the tail of distribution. Figure 2 depicts expected costs as a function of the number of diversified banks.  It is clear that the optimal fraction of diversified banks increases with $N ( =K)$. 


\section{Externality}

\subsection{Basic framework}

    Thus far, it has been assumed that a financial market consists of isolated banks only. In this section, I introduce interbank networks and see how the previous results will be altered. Introducing a network structure not only correlates the states of individual banks, but also differentiates each bank\rq{}s systemic importance. In the presence of interbank trading, a default of one bank could generate further defaults. A default cascade, or contagion, is more likely to occur when a more \lq\lq{}infective bank\rq\rq{} fails. Therefore, regulators should internalize the potential externality effects, the extent of which could differ from bank to bank. 


 Let us describe a model of the interconnected financial market here. A typical bank\rq{}s balance sheet is illustrated by Figure 3. It is assumed that the unit volume of interbank lending is fixed for all banks and that the composition of the balance sheet is common to all banks, which means that more interconnected banks are bigger in size. This commonality of balance sheet structure ensures that the probability of ``fundamental default\rq\rq{} is \textit{prima facie} identical across banks regardless of bank size. By fundamental default, I mean the failure of a bank caused purely by the loss of external assets.  In this model, the differences in the topological location in the network are reflected solely by the portfolio of interbank assets.

The asset side of bank $i$'s balance sheet consists of risky external assets, $a_{i}$, interbank assets, $l_{i}$, and a riskless asset, $b_i$. At the maximum, each bank may hold $K$ kinds of risky external assets and $N-1$ interbank assets, 
where $a_i = \sum_{k=1}^K a_{i,k}$ and $l_i =  \sum_{j \neq i}^N l_{i,j}$.  
The liability side of bank $i$'s balance sheet consists of interbank liability, $\bar{p}_{i}$, deposits, $d_i$, and
net worth, $w_i$. The balance sheet condition implies that $a_i + b_i + l_i$ = $\bar{p}_i + d_i + w_i$, $i = 1,\ldots , N$. 
 The size of the balance sheet is determined such that the relative proportions
 of $a_i$, $b_i$, and $l_i$ are common to all interconnected banks, given the unit size of interbank assets.
The portfolio of external risk assets, $(a_{i,1},\ldots ,a_{i,K})$, can vary from bank to bank. 

The amount of bank $i$'s borrowings from bank $j$ is expressed as $\pi_{ij}\bar{p}_i$, where $\pi_{ij}$ denotes the
relative weight of bank $i$'s borrowings from $j$, and thereby $\sum_{j\neq i} \pi_{ij}=1,\; i = 1,\ldots , N$. 
Let the $N$-by-$N$ matrix $\bm{\Pi}$ have $\pi_{ij}$ as its $(i,j)$-th element.
The diagonal elements are all zero because no banks lend to themselves. The amount of bank $i$'s total interbank assets,
$l_i$, can be given as $$ l_i  =  \sum_{j \neq i}^N l_{i,j} = \sum_{j\neq i}^N \pi_{ji}\bar{p}_j, $$ 
In vector form, $$ \bm{l} = \bm{\Pi} '\bm{\bar{p}}.$$

 Bank $i$ tries to pay back the full amount of interbank liability $\bar{p}_i$ after the values of external assets are realized, 
 but the funds available at that time are expressed as
 \begin{equation}
 \sum_{j\neq i}\pi_{ji}{p}_j + \tilde{a}_i + b_i-d_i,
 \end{equation} 
where $\tilde{a}$ denotes the ex-post value of external assets. It should be pointed out that deposits, $d_i$,
are reserved because deposits are senior to interbank assets. 
 I also impose three natural conditions: (a) limited liability; (b) the priority of debt claims; and (c) the proportionality of the amount of repayment after default (c.f., Eisenberg and Noe \cite{eisenberg01}).
The actual amount of repayment, $p_j$, is not necessarily equal to its face value, $\bar{p}_j$, since the sum of bank $j$'s loss of interbank and external assets might be greater than its net worth, $w_j$. 

 More precisely, the market-clearing vector, $\bm{p}^*$, is expressed as a fixed point of the map, $\bm{\Phi} : [\bm{0},\bm{\bar{p}}] \to [\bm{0},
\bm{\bar{p}}]$ defined as
 \begin{equation}\label{fixed} \bm{\Phi} (\bm{p}) = \bm{\bar{p}} \wedge (\bm{\Pi} ' \bm{p} +
\bm{\tilde{a}} + \bm{b} -\bm{d}) , 
\end{equation} 
where $\wedge$ denotes the meet operator, such that $\bm{ x} \wedge \bm{y} = (min(x_1, y_1),\ldots ,min(x_n, y_n)). $
Under the reasonable conditions (a), (b), and (c), Eisenberg and Noe \cite{eisenberg01} prove that a fixed point of (\ref{fixed}) exists and is unique
 if the financial network is strongly connected and $\sum_i (\tilde{a}_i + b_i -d_i ) > 0$. They also show that the unique fixed point is the solution to the following linear program:
$$ \max_{\bm{p}\in[\bm{0},\bm{\bar{p}}]} \: f(\bm{p}) \;\: {\rm s.t.}\;\; \bm{p} \leq \bm{\Pi}\rq{}\bm{p} + \bm{\tilde{a}} + \bm{b} -\bm{d} 
$$
for any strictly increasing function $f(\cdot ): \mathbb{R^{\rm N}} \to \mathbb{R}.$

The simulation procedure is as follows. First, the topology of the network structure is given. Given the
unit size of interbank assets, the balance sheet size and thereby the volume of each component are automatically determined. In order to
apply the Eisenberg-Noe theorem, it is assumed that any two banks that are connected by an edge lend to 
each other, i.e., edges are bidirectional. This is the simplest assumption that ensures that any interconnected networks are also strongly interconnected. 
Second, random returns of $K$ external risk assets are drawn and fundamental defaults are observed.
 Third, given the funds available for the repayment of interbank loans, $\bm{\tilde{a}} + \bm{b} -\bm{d}$, the unique market-clearing vector,
$\bm{p}^*$, is calculated using the Eisenberg-Noe algorithm.  Because a fundamental default may cause contagion,
the number of default banks would increase at this stage, depending on the structure of the network. Such additional defaults are called contagious defaults. It is assumed that if  $\sum_i (\tilde{a}_i + b_i -d_i ) \leq 0$, then all banks go bankrupt. 

\begin{table}
\caption{Baseline parameters}
\arrayrulewidth = 2pt
  \begin{tabular}{ccccc} \hline 
  cost parameter & capital/asset & external asset/capital & I.B. asset/capital & I.B. liability/capital  \\ \hline 
   $ s$ & $\frac{w_i}{a_i+b_i+l_i}$ & $a_i/w_i$ & $l_i/w_i$ & $\bar{p_i}/w_i$ \\ 
4 &.1& 5& 4 & 4  \\ \hline 
  \end{tabular}
\end{table}

  The baseline parameters are summarized in Table 1. The return of each external asset is assumed to follow an $i.i.d.$ normal distribution with mean zero. The variance of each asset return is determined such that the probability of fundamental default is equal to $p$ if the bank holds a single external asset.

\subsection{Independent external assets}

Suppose that $N=5$ and $K=3$. Figure 4 illustrates how the expected costs depend on the distance between banks, $D$, and the distance from the optimal
portfolio, $G$, which are defined as
 \begin{eqnarray} D &\equiv &\frac{1}{2N(N-1)}\sum_i \sum_j \sum_k \left| a_{i,k} -
a_{j,k} \right|, \\ G &\equiv &\frac{1}{N}\sum_k\left| \sum_i (a_{i,k} - 1/K) \right| . 
\end{eqnarray} 
$D =G=0$ corresponds to the case of full diversification. Beale et al. \cite{beale11} demonstrate that when $s > 1$, the social costs tend to be reduced by increasing $D$. This is because the greater the distance between banks, the lower the probability of simultaneous defaults. This is reconfirmed by the simulation without an interbank network (Figure 4a). However, it turns out that this outcome does not hold true once the possibility of
contagion is taken into account (Figures 4b, 4c, 4d). The introduction of externality moves the optimal combination of $(D, G)$ toward the origin. It is interesting to note that the optimal value of $D$ reaches zero as all banks become interconnected. I found that this phenomenon remains unchanged even when the relative amount of interbank assets, $l_i/w_i$, is reduced to 1. A simple rationale is that as banks become interconnected, the financial system takes the form of one ``big bank\rq{}\rq{}. Thus, full diversification becomes desirable because it attains the minimum probability of default for the ``big bank\rq\rq{} by analogy with a single individual bank. This implies that the stability of an interconnected financial system will be best achieved by micro-prudential policies. One may think that if the probability of fundamental default is very small,
then diversification might not be desirable. It is true that a smaller possibility of individual failure will make a default cascade less likely, but
it also reduces simultaneity risk. The latter effect is likely to dominate the former, suggesting that the advantage of diversification is greater than that of diversity.       


  To examine in greater detail how the topology of the network structure affects financial fragility, Figure 5 illustrates eight patterns of five-bank network topologies. Although a five-bank network is unrealistically small, a single ``bank\rq{}\rq{} in this network could also be viewed as a group of financial institutions that have an identical portfolio. Each network structure is named after its degree distribution. For example, structure (b) 2-1 contains nodes that have degree 1 or 2. Figure 6 is a visualization of the likelihood of contagious defaults, where each bank is assumed to hold one bank-specific external asset.
 The number of dots in the $(i,j)$-th element indicates the expected number of defaults of bank $j$ per 1000-time single defaults of bank $i$. Based on this information,
 Figure 7 shows the extents of  ``infectivity'' and ``susceptibility''. By the infectivity of bank $i$, I mean the average number of contagious defaults generated by a single fundamental default of bank $i$. The susceptibility of bank $i$ represents the ratio of the number of times that bank $i$ suffers from infectious defaults to the number of bank $i$\rq{}s fundamental defaults. In order to see the topological importance of each node in the network, PageRank (Brin and Page \cite{brin98}) $y_i$ is also shown in the figure, where
$$  y_i = \alpha \sum_j A_{ij}\frac{y_j}{k_j^{out}} + \beta  .
$$
$A_{ij}$ is the $(i,j)$-th element of the adjacency matrix, which takes a value of one if  $\pi_{ij} > 0$ and zero otherwise. Notice that $A$ is a symmetric matrix in this model. $k_j^{out}$ is the out-degree of node $j$. The parameter values are $\alpha =.85$ and $\beta = 1$.   

It turns out that a bank\rq{}s infectivity is very closely correlated with its PageRank. In fact, this property can be obtained only if PageRank is used as the centrality measure. Other centrality measures, such as the betweenness centrality, the Katz centrality, the closeness centrality, and the eigenvector centrality, are not necessarily correlated positively with infectivity. As for network (b), for instance, only PageRank can correctly report that nodes 2 and 3 are equally the most influential. Other centrality measures predict that node 1 is the most important. The reason for this difference is that PageRank assigns a high centrality to a node that is pointed to by a node that has a small number of outgoing edges. Bank 1 in network (b) is located at the center of the network, but its counterparties, banks 2 and 3, each have two edges. On the other hand, banks 2 and 3 are each linked to a bank that has only one edge, banks 4 and 5, respectively. In the latter case, the failure of bank 2 (3) will cause the failure of bank 4 (5) with high probability. Therefore, a bank with higher PageRank becomes more infective because a higher-ranking bank collects a larger fraction of credit from each of its counterparties.

 While PageRank is a good indicator of the topological importance of individual banks, it is also useful to construct a measure of the topological fragility of a financial system as a whole. Figure 8a illustrates the expected costs under full diversity and network entropy, $S_N$, defined as\footnote{Wilhelm and Hollunder \cite{wilhelm07} propose various information-theoretic measures for network characterization. Sato \cite{sato10} employs entropy measures to describe the state of foreign exchange markets.}
\begin{equation}
   S_N = -\sum_{j=1}^N p_j\log{p_j}, \:\mathrm{where} \:\: p_j = \frac{\# \mbox{ of bank $j$\rq{}s links}}{\# \mbox{ of total links in the network} }.
\end{equation}
Figure 8b shows the relationship between the expected costs and the Herfindahl-Hirschman Index (HHI), where
\begin{equation}
   HHI_N = \sum_{j=1}^N p_j^2.
\end{equation}
It turns out that $S_N$ and $HHI_N$ have clear negative and positive relationships with the expected costs under full diversity, respectively. This implies that the topological riskiness intrinsic to a financial network is well captured by the degree of market concentration. If the number of links is quite similar among banks, which means that the size of the banks is similar, then the financial system is relatively stable. In contrast, if a small number of banks own a large fraction of the total links in the network, i.e., if there are a small number of giant banks, then the financial system becomes fragile. Network (d) is the most notable example. This property is unchanged if PageRank is used to define $p_j$ instead of degree. Notice that a simple measure of connectivity, such as the average degree, does not provide accurate information on the level of systemic risk.

  Figure 9 illustrates the optimal allocation of external assets when $N=5$. If there is no interbank trading, then the optimal situation for society is that any one bank diversifies its portfolio and the other four have independent bank-specific assets. In contrast, if there is interbank trading and all banks become interconnected, then the socially optimal state is that all banks diversify their portfolios. This outcome holds true regardless of the topology of the financial network. This implies that the previous argument for the case of $N=5$ and $K=3$ continues to hold for a more general structure of a financial network. It could be said that micro-prudential policies will add up to a macro-prudential policy as long as the returns of original assets are independent. 

\subsection{Correlated external assets}

Although, as PageRank suggested, the topological importance of each bank differs substantially, the above results indicate that the most desirable state for an interconnected financial market is that every bank fully diversifies its portfolio. Recall that a presumption for this argument is that the returns of external assets are independent of each other. In the following, I consider a more general case in which external assets may be correlated.  

  Consider again the case of $N=5$. The set of external assets, $\{ a_i\}$, available for banks is now given as 
$$\{a_1,a_2,a_3,a_4,a_5,a_6 \}=\{ a_1, -\rho a_1\!+\! (1\!-\!\rho )\hat{a}_2, a_3, \rho a_3\! +\! (1\!-\!\rho )\hat{a}_4, a_5, \frac{1}{5}\sum_{i=1}^5 a_i \},$$ 
where $\rho \in [0,1]$, and the return of each element of $(a_1,\hat{a}_2,a_3,\hat{a}_4,a_5)$ follows an independent normal distribution. Notice that if $\rho > 0$, assets 1 and 2 are negatively correlated while assets 3 and 4 are positively correlated. Asset 5 is independent of assets 1-4.    
  For the variance of each asset not to be dependent on $\rho$, it is assumed that $\hat{\sigma}^2 = \frac{1+\rho}{1-\rho}\sigma^2$, where $\sigma$ and $\hat{\sigma}$ denote the standard deviations of $a_i$ and $\hat{a}_i$, respectively. If $\hat{\sigma}$ is so defined, then the variance of the diversified asset $a_6$, $var[\frac{1}{5}\sum_{i=1}^5 a_i]$, is also independent of the value of $\rho$.  
Thus, the set of external assets used here differs from the previous one only in terms of the presence of correlations. It is assumed that banks are allowed to hold one of the six external assets.
 The number of asset portfolio patterns to be examined is $6^N = 7776$ for each network structure. Random returns of each asset are drawn 5000 times.   


 Figure 10 shows the optimal allocation of external assets. There are two notable features: first, neither of the positively correlated assets, $a_3$ and $a_4$, nor the independent asset, $a_5$, is used in interconnected networks, while both of the negatively correlated assets, $a_1$ and $a_2$, are used in every network structure. This reflects the fact that the two negatively correlated assets are advantageous in that they generate little possibility of simultaneous defaults.   

Second, the topologically most influential node, in terms of both infectivity and PageRank, is not necessarily assigned the diversified asset, $a_6$. In networks (e), (f) and (g), for example, the infectivity measure and PageRank both indicate that node 1 is the most important, but asset 1 is assigned to node 1 in all cases. This might be somewhat counterintuitive because the diversified asset ensures the lowest probability of fundamental default. Note that if $p = .2$, then the probability of fundamental default by a bank that holds the diversified asset, $a_6$, is approximately $.03$. Thus, the likelihood of a fundamental default by a bank with an asset other than $a_6$ is about 6.7 times higher than that by a diversified bank. This implies that in certain network structures, something is more important than infectivity from the standpoint of systemic stability. 

The reason why the most infective bank need not be the safest is twofold: first, collective fundamental defaults would be more influential than single defaults. To see this, Figure 11 shows the contributions of various combinations of collective defaults to the total expected costs. It turns out that in all networks but (d), a single default of node 1 is not the main contributor to systemic risk.  Rather, double and triple defaults are systemically more important than a single default of node 1. This suggests that double and triple defaults be avoided to reduce systemic risk, which makes it desirable to have the banks\rq{} balance sheets move in counteractive ways. As for network (d), however, node 1 is so infective that it should be protected with the highest priority. In this sense, node 1 of network (d) could be viewed as a ``super spreader\rq{}\rq{}. Second, a highly connected node is intrinsically safer than a poorly connected one. This can be confirmed by Figure 7, which indicates that a node with higher infectivity tends to exhibit lower susceptibility. This is because, given the external asset portfolio, interbank assets become more diversified as the number of links increases. Such a higher extent of diversification of interbank assets creates room for external assets to be less diversified.


 Figure 12 shows how the optimal allocation of assets varies with the degree of convexity of the cost function. Only networks (b) and (e) are shown, but essential results hold true also for other network structures. There are two notable phenomena. First, the number of banks that hold asset 6 decreases with $s$. When $s$ is equal to or larger than 8, it is optimal that only one bank holds the diversified asset. This is because simultaneity risk becomes more severe as $s$ increases. At this point, one may wonder why holding asset 1 or 2 would be systemically less costly than holding asset 6.  To simplify  the situation, suppose that there are three isolated banks and the regulator considers two types of asset allocation: (a) $(a_1, a_2, a_2)$ and (b) $(a_1, a_6, a_6)$. It is assumed that the probabilities of fundamental default are $p$ for holding $a_1$ or $a_2$ and $q\: (< p)$ for holding $a_6$. Since $a_1$ and $a_2$ are negatively correlated, suppose that there is no risk of double default for any combination of two banks such  that one has $a_1$ and the other has $a_2$. $a_1$ and $a_2$ are assumed to be uncorrelated with $a_6$. 
 For patterns (a) and (b), the expected costs can be respectively calculated as  
\begin{eqnarray}
C_a &=& p + p \cdot 2^s, \\
C_b &=& p(1-q) + q(1-p) \cdot 2^s + pq\cdot 3^s. 
\end{eqnarray}
Note that there is no risk of triple defaults in the case of (a). It follows that $C_b \geq C_a$ if and only if
 \begin{equation}\label{cacb}
3^s \geq 1+\left( 1+ \frac{p-q}{pq} \right) 2^s
\end{equation}
Since $p > q$, (\ref{cacb}) does not hold if $s=1$. However, (\ref{cacb}) surely holds for values of $s$ greater than or equal to a certain threshold $s^*$ such that $ 3^{s^*} = 1+\left( 1+ \frac{p-q}{pq} \right) 2^{s^*}$.  This simple example clearly show that the benefit of a counteractive portfolio structure is increasing in $s$. Although diversified assets increase the soundness of individual balance sheets, this does not necessarily lead to a reduction of systemic risk if the cost of simultaneous defaults is sufficiently large (i.e., $s>s^*$). For sufficiently large values of $s$, the risk of triple defaults becomes significant even though its likelihood is quite low. Thus, introducing a counteractive portfolio structure becomes more desirable as a larger size of collective defaults begins to contribute to systemic risk.   

 The second point to be noted is that the safest node would differ as the value of $s$ varies. As shown in the figure, node 1 is assigned asset 1 when $s=4$ but asset 6 when $s=15$. Recall that node 1 is not the most infective bank in network (b), so that it is not correct to conjecture that the most infective node should be the safest when $s$ is sufficiently large.  A possible explanation is that node 1 should act as a firewall, because it is located at the center of the network. For example, if $s=8$, there is a possibility of triple fundamental defaults by banks that each have $(a_1,a_1,a_6)$ or $(a_2,a_2,a_6)$. Although there is little possibility of quadruple fundamental default, it is highly likely that node 1 would suffer from infectious default if both of its counterparties fail. In contrast, in the optimal portfolio structure under $s=15$, there is little possibility that both of bank 1\rq{}s counter parties fail by fundamental default. Therefore, if large-scale collective defaults are highly contributive to the expected costs, then it is systemically beneficial to protect the node that would effectively act as a firewall. In the terminology of network theory, node 1 is called a \lq\lq{}cut point\rq{}\rq{} in the sense that removing node 1 will separate the original network into two isolated components. From this viewpoint, protecting the cut point will prevent a collective default of one component from transmitting to the other. 


 Figure 13 illustrates how the introduction of counteractive portfolio structures reduces expected costs. It shows that the relative benefit of introducing negatively correlated assets varies with the network structure.  An important fact is that the relative costs under optimal asset allocation are not necessarily correlated with those under full diversity. Recall that the case of full diversity exhibits pure effects of network topology on systemic risk. Although the relative extent of systemic risk under full diversity has a negative correlation with network entropy, such a correlation disappears once a system-wide portfolio structure is optimized. This implies that the fragility of a financial system is totally unknown until asset allocation is taken into account. In other words, optimization of asset allocation could eliminate the differences in systemic risk stemming from differences in network topology.

\section{Discussion and conclusion}

 The recent literature addresses systemic risk mainly from two different and independent viewpoints:  asset allocation and network structure.  A recent study by Beale et al. \cite{beale11} is an example of the former.  It is important to note, however, that their argument misses a crucial aspect of actual financial systems: interconnectivity. 
On the other hand, many studies, such as Gai and Kapadia \cite{gai10}, Gai et al. \cite{gai11}, and Hurd and Gleeson \cite{hurd11}, focus on investigating the relationship between network structure and systemic risk, employing a methodology developed in network theory. In these studies, the role of asset commonality is missing. The model of this paper incorporates both viewpoints, considering a situation in which the extent of systemic risk depends not only on how banks choose their asset portfolio, but also on how banks are interconnected with each other. 

 The optimal allocation of external assets that minimizes systemic risk naturally depends on the extent of both simultaneity risk and externality. The analysis reveals that the use of negatively correlated assets will be quite effective in reducing systemic risk by lowering the probability of collective fundamental defaults.  An important result is that the most infective bank, or the topologically most influential bank indicated by PageRank, need not always be the safest. Under certain circumstances, the most infective node is required to act as a firewall to prevent large-scale collective defaults. 
From this point of view, it appears that the widely used term ``systemically important financial institution (SIFI)\rq{}\rq{} is harder to define than is usually thought. Recently, Caccioli et al. \cite{caccioli12} also argue that increasing the capital buffer of most highly connected banks may not affect systemic risk.

 The optimization of system-wide asset allocation also reveals that the relative fragility of a financial network structure is unknown until asset allocation is taken into account. This is quite important because systemic risk tends to be considered directly related to network topology. However, differences in the amount of systemic risk due to differences in network topology could be eliminated to a large extent by appropriately regulating the asset holdings of banks. 

 In recent years, there has been wide agreement that countercyclical capital buffers are needed to reduce economic fluctuations. This is based on the idea that the previously employed simple capital requirement could amplify business cycles due to the procyclicality of capital caused by the regulation itself. Roughly speaking, this paper shows that these arguments can also be applied in a crossectional dimension. In order to reduce the risk that the balance sheets of a large number of banks fluctuate in sync, a certain fraction of banks should have balance sheets that fluctuate counteractively. This will lower the level of simultaneity risk, ensuring bank diversity.

 Some issues need to be addressed in future research. First, the validity of certain characteristic measures of network structures, such as modularity and assortativity, in determining asset allocation should be explored in more detail. To do this, a larger-scale financial network needs to be considered, but computation will be highly time-consuming. Second, the interconnectivity of banks needs to be treated in a more general form. In this model, it is assumed that banks are strongly connected in order to apply the Eisenberg-Noe algorithm in calculating market-clearing vectors. However, the actual interbank market is not necessarily strongly connected, and there are many one-directional edges and weakly connected components (S\"oramaki et al. \cite{soramaki07}, Iori et al. \cite{iori08}, Imakubo and Soejima \cite{imakubo10}). Another algorithm needs to be employed to explore such a complex financial network.
 
\section*{Acknowledgments}
 I would like to thank Charls Kahn, Takashi Kamihigashi, Wataru Takahashi, Takayuki Tsuruga, seminar participants at Kobe University and the Bank of Japan and two anonymous referees for their helpful comments. I also thank Kohei Hasui for the excellent research assistance. Financial support from the Japan Securities Scholarship Foundation and KAKENHI 25780203 and 24243044 are gratefully acknowledged.

\newpage
%
%
\begin{figure}
\center
\includegraphics[width=8cm,clip]{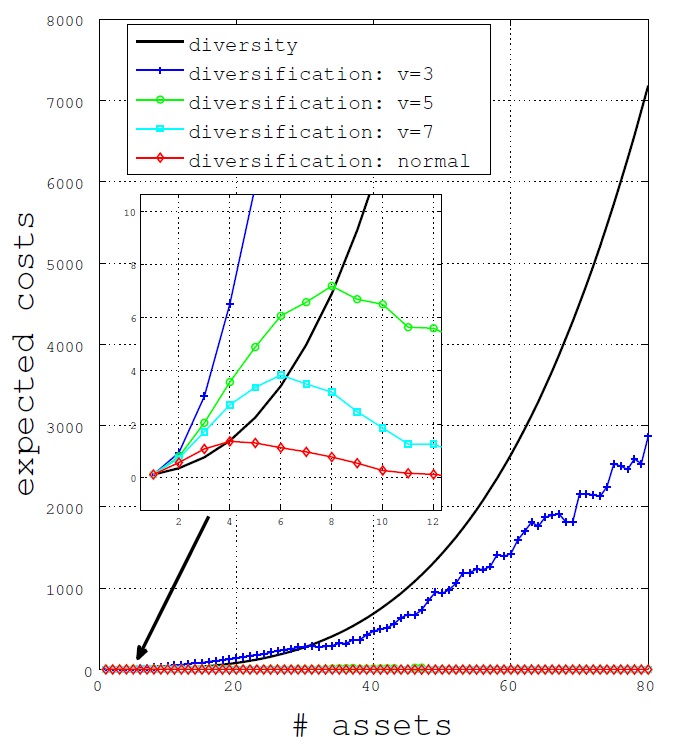}
\caption{\setlength{\baselineskip}{14pt}Expected costs under full diversity and full diversification: asset returns obey Student t-distribution with degree of freedom $v$. There is no interbank network. The number of assets is assumed to be equal to the number of banks. Random returns are drawn 200,000 times for each asset. $p=.1$ and $s=4$.} 
\end{figure}
%
%
\begin{figure}
\center
\includegraphics[width=16cm,clip]{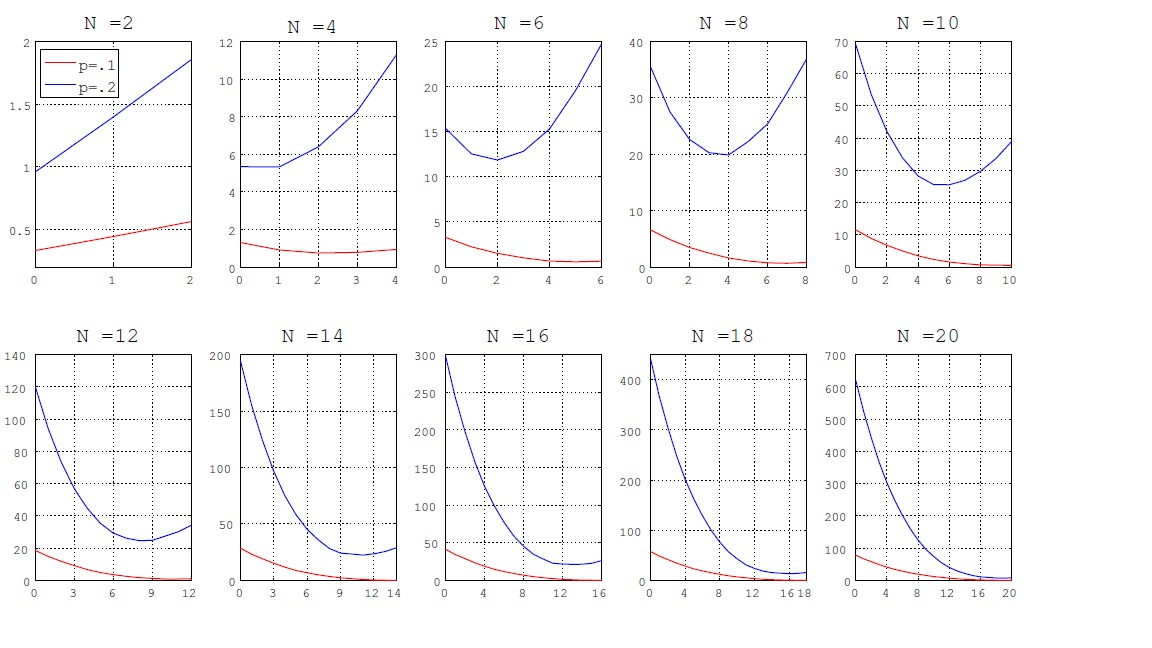}
\caption{\setlength{\baselineskip}{14pt} Expected costs with varying numbers of diversified banks. Horizontal axis: number of diversified banks. Vertical axis: expected costs. Banks other than diversified ones have bank-specific assets. Each asset return is assumed to follow an independent standard normal distribution. There is no interbank network. Random returns are generated 20,000 times for each asset. $s=4.$}
\end{figure}
%
%
\begin{figure}
\begin{center}
\includegraphics[width=8cm,clip]{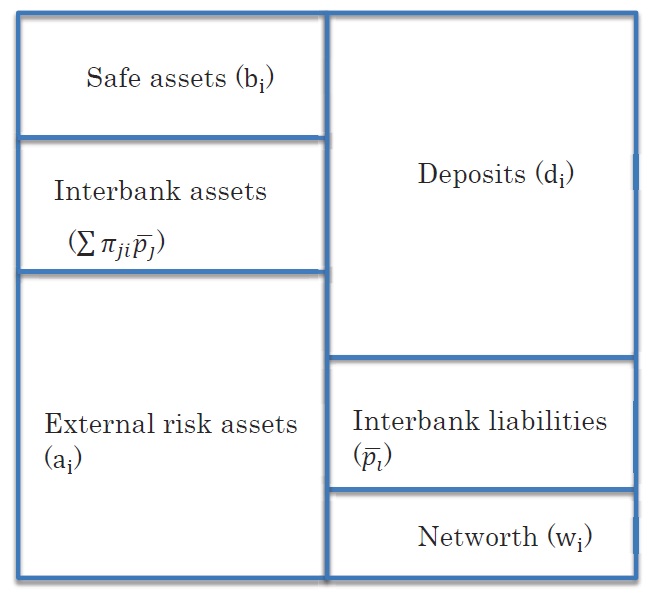}
\end{center}
\caption{A typical bank\rq{}s balance sheet.} 
\end{figure}
\begin{figure}
\begin{center}
\includegraphics[width=13cm,clip]{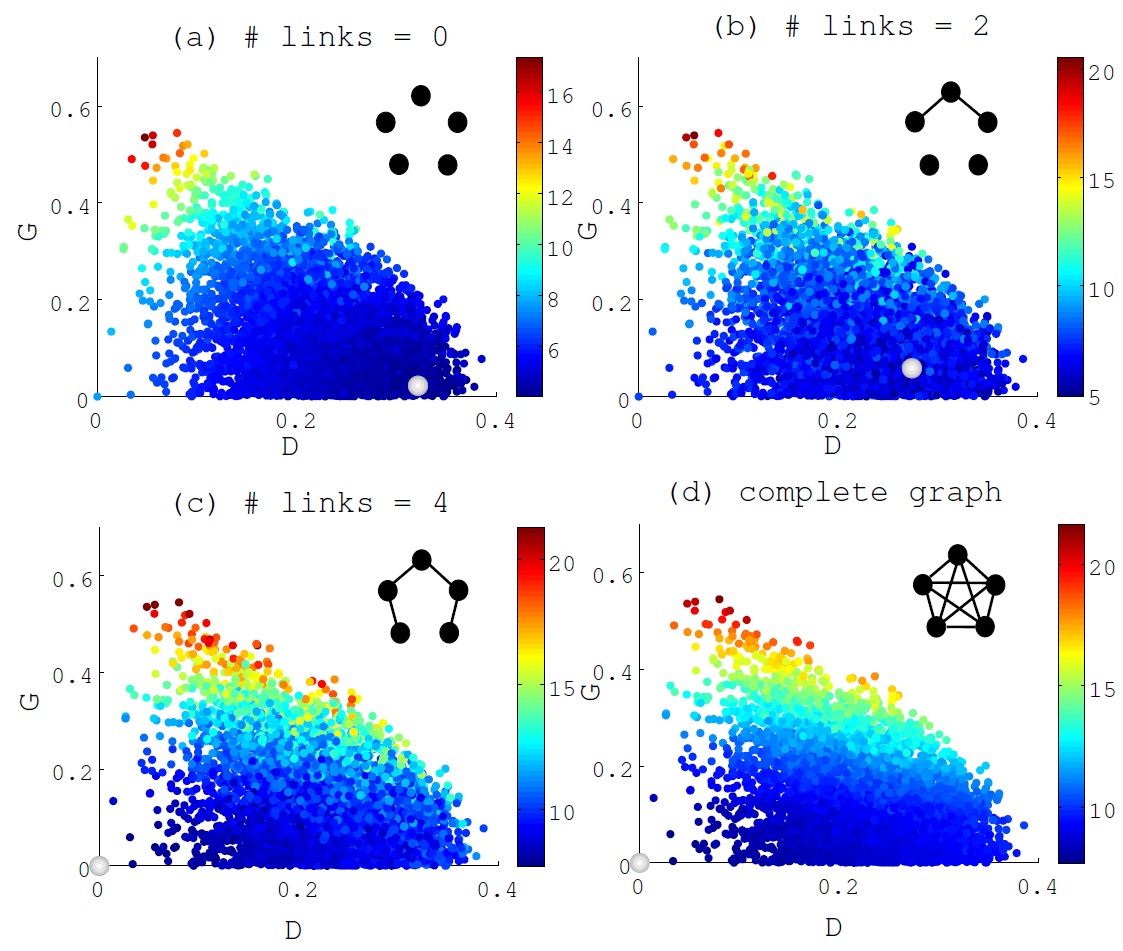}
\caption{\setlength{\baselineskip}{14pt}Expected costs with various combinations of $D$ and $G$ (defined by eqs.(6) and (7), respectively). Combinations of $(D,G)$ that attain relatively low (high) expected costs are indicated by cold (warm) colors. 
$N$=5, $K$=3 and $p$=.1. (a) There is no interbank network. (b) and (c) The numbers of links are 2 and 4, respectively. (d) Complete graph. 
Random returns are drawn 20,000 times for each of 5,000 portfolio patterns. The combination of $(D,G)$ that minimizes the expected cost is indicated by a gray sphere.} 
\end{center}
\end{figure}
%
%
\begin{figure}
\center
\includegraphics[width=13.5cm,clip]{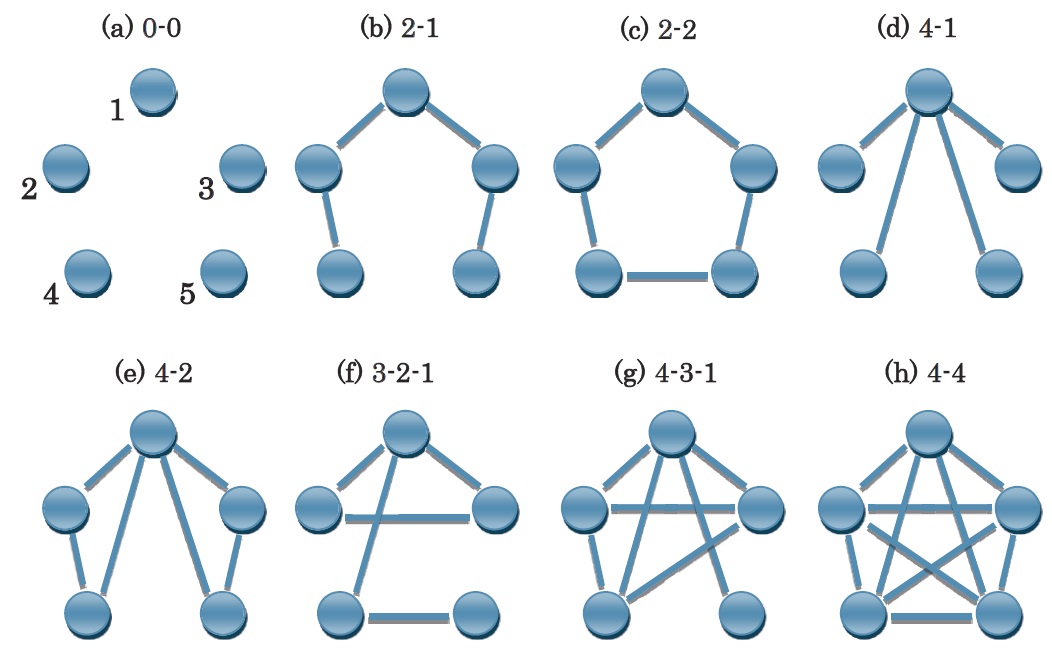}
\caption{\setlength{\baselineskip}{14pt}Eight patterns of interbank networks. Each network structure is named after its degree distribution.} 
\end{figure}
%
%
\begin{figure}
\center
\includegraphics[width=15.5cm,clip]{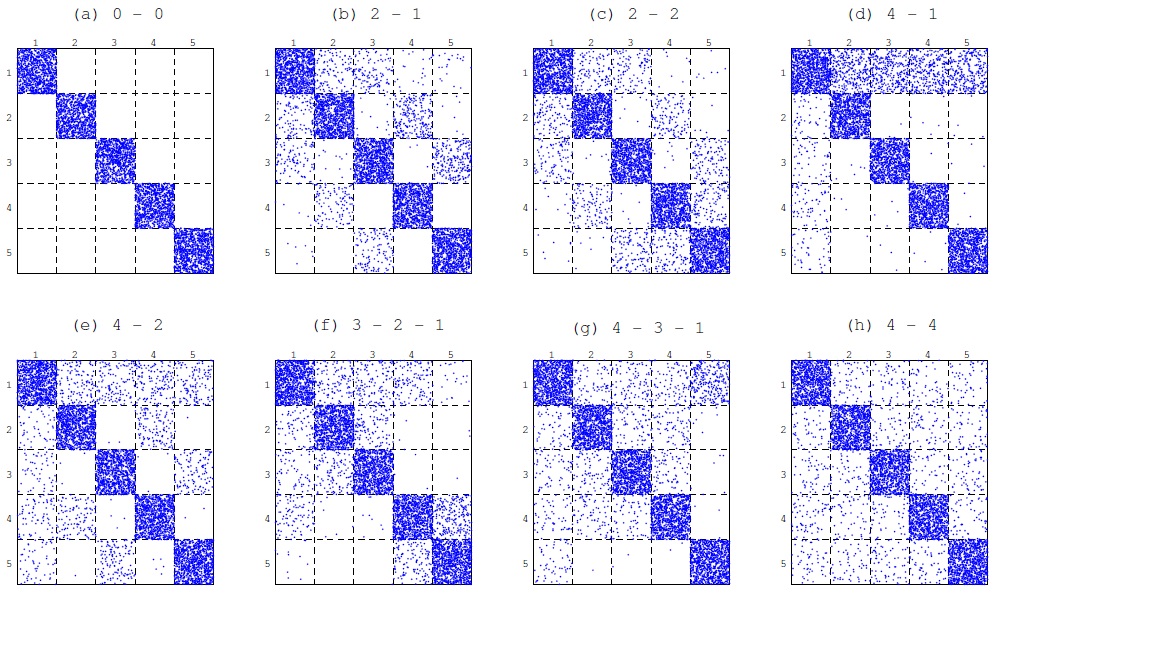}
\caption{\setlength{\baselineskip}{14pt}Visualization of the likelihood of contagious defaults under the eight patterns of network structure shown in Fig.5. The number of dots in the $(i,j)$-th element indicates the expected number of defaults of bank $j$ per 1000-time single defaults of bank $i$. Full diversity is assumed. $p=.2$.} 
\end{figure}
%
%
\begin{figure}
\center
\includegraphics[width=15.5cm,clip]{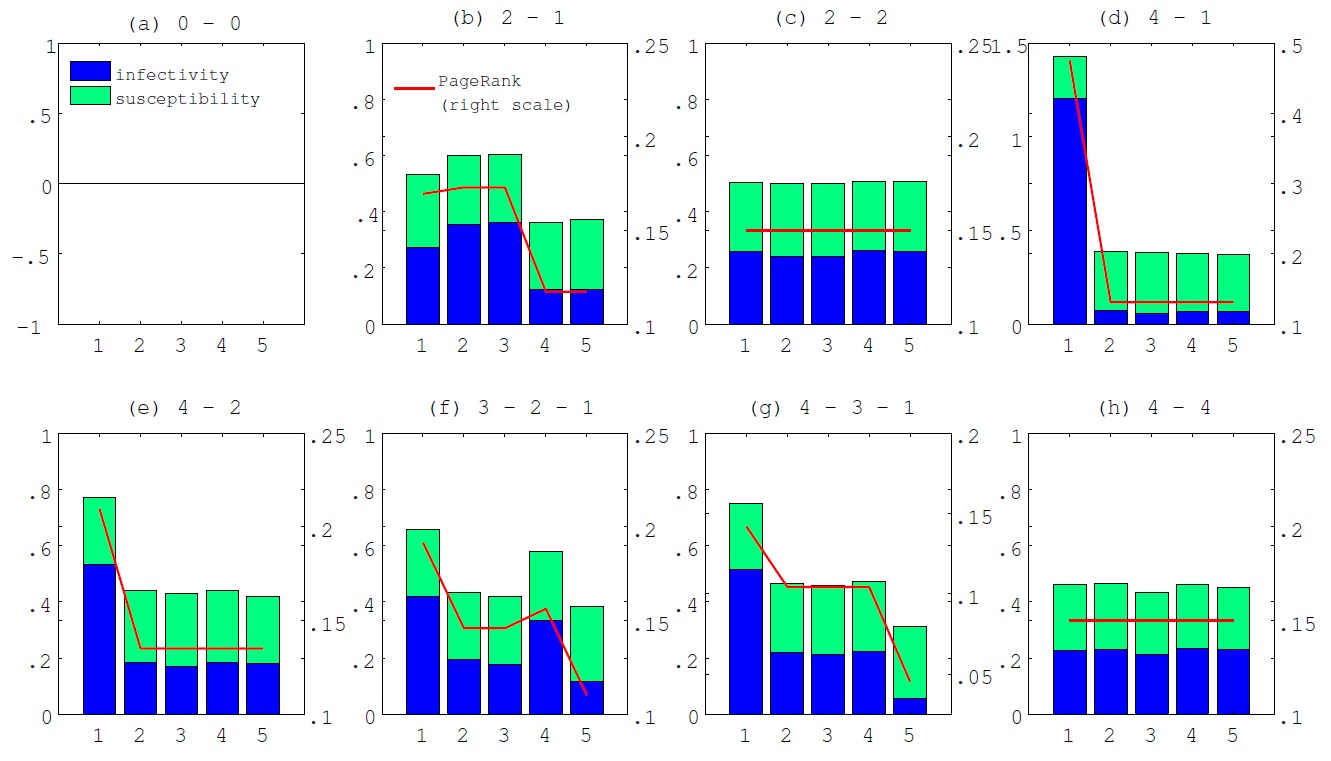}
\caption{\setlength{\baselineskip}{14pt}Infectivity, susceptibility and PageRank  under the eight patterns of network structure shown in Fig.5. Horizontal axis: bank index. Vertical axis (left): infectivity (blue) and susceptibility (green). Vertical axis (right): PageRank (red solid line).  
The infectivity of bank $i$ is defined as the average number of contagious defaults generated by a single fundamental default of bank $i$. The susceptibility of bank $i$ represents the ratio of the number of times that bank $i$ suffers from infectious defaults to the number of bank $i$\rq{}s fundamental defaults. Full diversity is assumed. $p=.2$.} 
\end{figure}
%
%
\begin{figure}
\center
\includegraphics[width=15.5cm,clip]{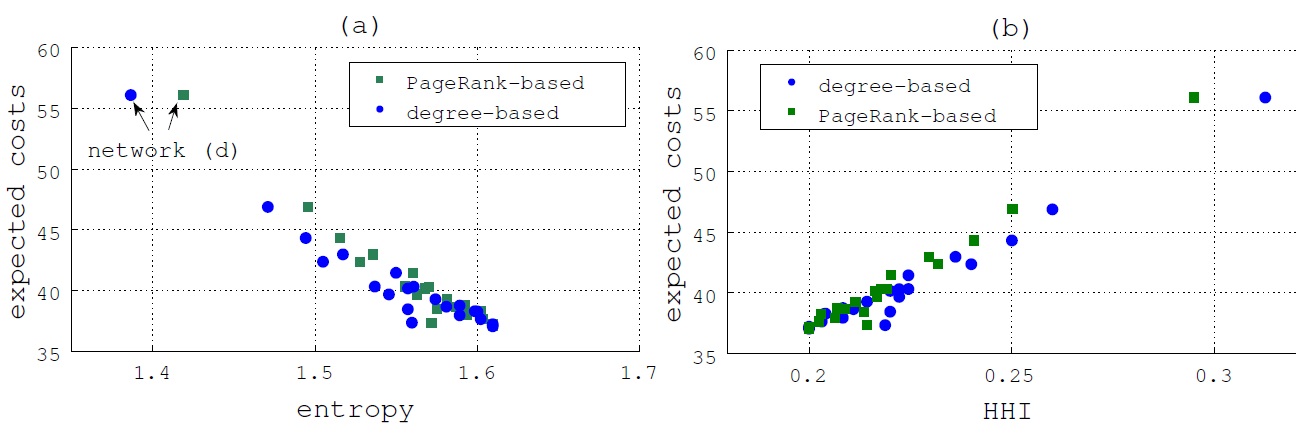}
\caption{\setlength{\baselineskip}{14pt}Expected costs under full diversity.  All the possible patterns of 5-bank interconnected network topologies are considered (21 patterns in total). $p=.2$. (a) Negative relationship between entropy and expected costs. Green square: entropy is based on PageRank. Blue circle: entropy is based on node degree.  (b) Positive relationship between the Herfindahl-Hirschman Index (HHI) and expected costs.} 
\end{figure}
%
%
\begin{figure}
\center
\includegraphics[width=13.5cm,clip]{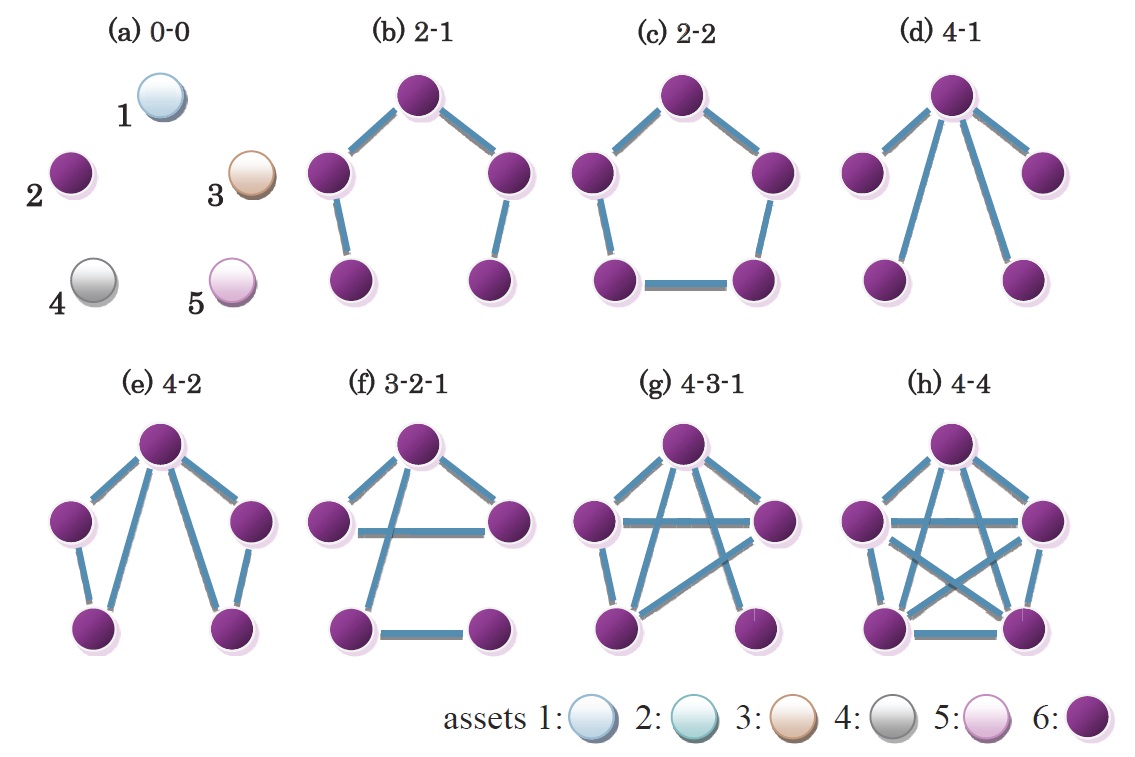}
\caption{\setlength{\baselineskip}{14pt}The optimal allocation of external assets for each topology. Assets 1-5 are bank-specific assets whose returns are independent ($\rho = 0$).  Asset 6 (dark purple circle) is the diversified asset, defined as the average value of assets 1-5. $p=.2$.} 
\end{figure}
%
%
\begin{figure}
\center
\includegraphics[width=13cm,clip]{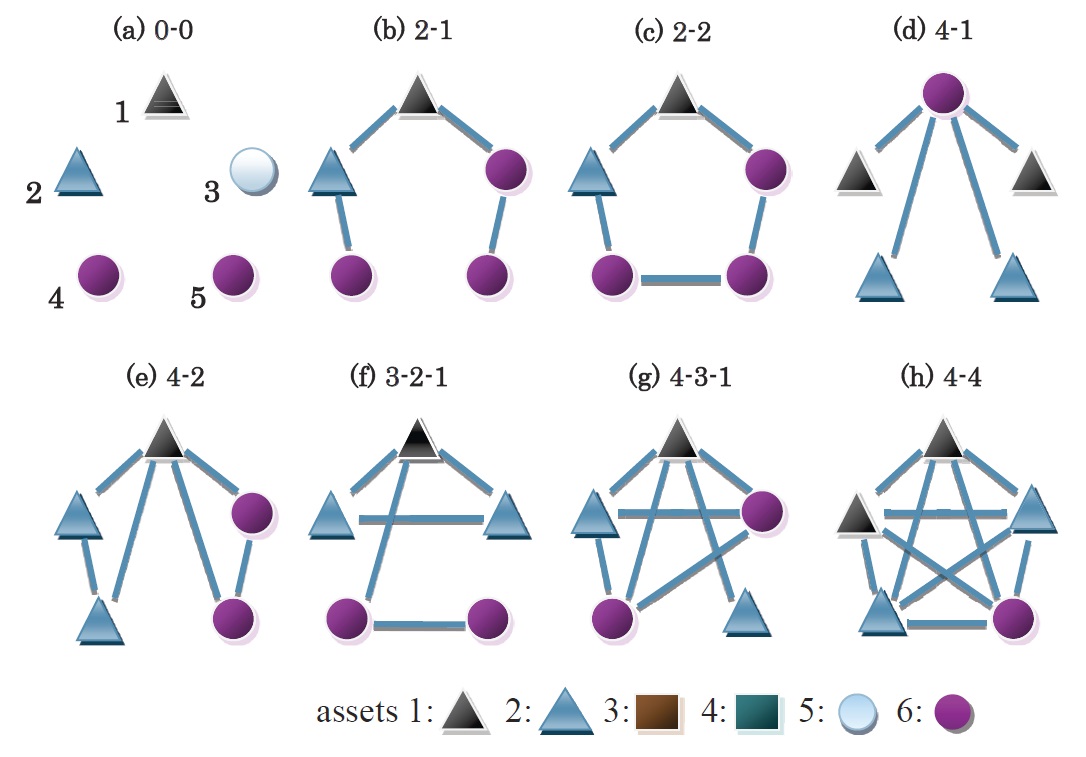}
\caption{\setlength{\baselineskip}{14pt} The optimal allocation of external assets for each topology. Assets 1 and 2 (black and light-blue triangles) have a negative correlation, while assets 3 and 4 (brown and blue squares) have a positive correlation. Asset 5 (white circle) is independent of assets 1-4. Asset 6 (purple circle) is the diversified asset, defined as the average value of assets 1-5.  $\rho =.8$ and $p=.2$.} 
\end{figure}
%
%
\begin{figure}
\center
\includegraphics[width=15.5cm,clip]{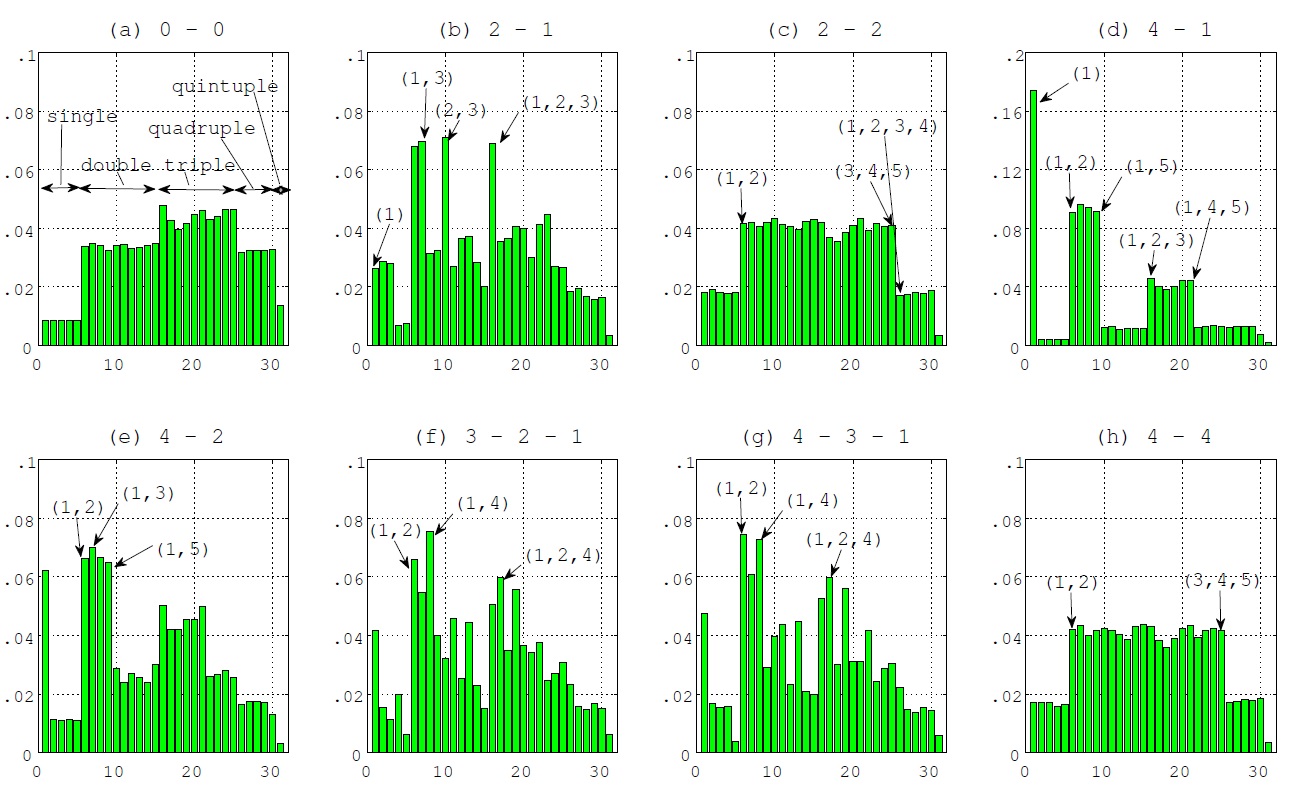}
\caption{\setlength{\baselineskip}{14pt} Contributions of various size of collective defaults to the total expected costs. Horizontal axis: index of collective defaults: \{1,2, ..., 31\} = \{(1), (2), (3), (4), (5), (1,2), (1,3), (1,4), (1,5), (2,3), (2,4), (2,5), (3,4), (3,5), (4,5), (1,2,3), (1,2,4), (1,2,5), (1,3,4), (1,3,5), (1,4,5), (2,3,4), (2,3,5), (2,4,5), (3,4,5), (1,2,3,4), (1,2,3,5), (1,2,4,5), (1,3,4,5), (2,3,4,5), (1,2,3,4,5)\}. Numbers in parentheses are indexes of banks that default simultaneously. Vertical axis: percentage of the cost of the corresponding collective default to the total expected costs. Full diversity is assumed. $\rho =0$ and $p=.2$.}
\end{figure}

\begin{figure}
\center
\includegraphics[width=12cm,clip]{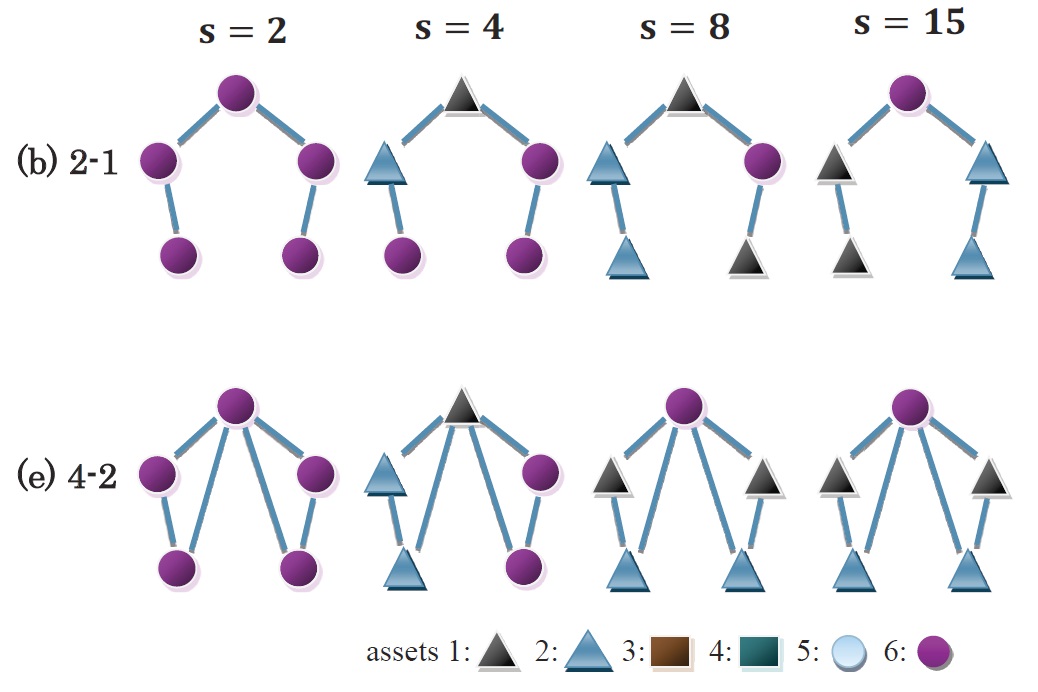}
\caption{\setlength{\baselineskip}{14pt}The optimal allocation of external assets with different values of $s$ (i.e., concavity of the cost function). Assets 1 and 2 (black and light-blue triangles) have a negative correlation, while assets 3 and 4 (brown and blue squares) have a positive correlation. Asset 5 (white circle) is independent of assets 1-4. Asset 6 (purple circle) is the diversified asset, defined as the average value of assets 1-5. $\rho =.8$ and $p=.2$.} 
\end{figure}

\begin{figure}
\center
\includegraphics[width=15cm,clip]{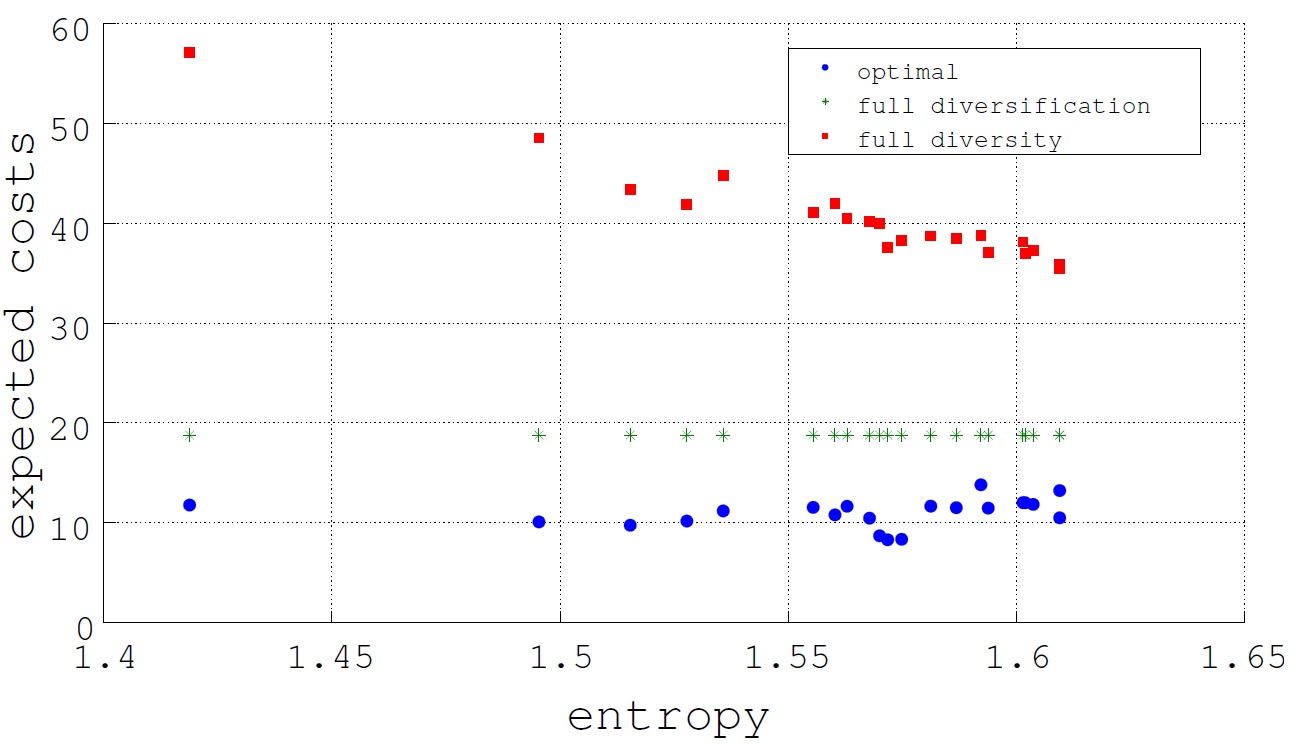}
\caption{\setlength{\baselineskip}{14pt}The relationship between the expected cost and the entropy based on PageRank. All the possible patterns of 5-bank interconnected network topologies are considered. Blue solid circles: expected costs under optimal allocation. Green stars: expected costs under full diversification. Red squares: expected costs under full diversity. $p=.2$. In deriving the optimal asset allocations, $\rho$ is set at .8.
The figure shows how the introduction of negatively correlated assets would reduce systemic risk.} 
\end{figure}


\begin{thebibliography}{99}

\bibitem{beale11}
N.~Beale, D.G. Rand, H.~Battey, K.~Croxson, R.M. May, and M.A. Nowak,
\newblock {P. Natl. Acad. Sci. USA}
  {\bf 108}, 12647 (2011)

\bibitem{upper11}
C.~Upper,
\newblock {J. Financ. Stability}. {\bf7}, 111 (2011)

\bibitem{may10}
R.M. May and N.~Arinaminpathy,
\newblock {J. R. Soc. Interface.} {\bf 7}, 823 (2010)

\bibitem{lenzu12}
S.~Lenzu and G.~Tedeschi,
\newblock {Physica. A} {\bf 391}, 4331 (2012)

\bibitem{gai10}
P.~Gai and S.~Kapadia,
\newblock {P. Roy. Soc. A} {\bf 466}, 2401 (2010)

\bibitem{gai11}
P.~Gai, A.~Haldane, and S.~Kapadia,
\newblock {J. Monetary Econ.} {\bf 58}, 453 (2011)

\bibitem{nier07}
E.~Nier, J.~Yang, T.~Yorulmazer, and A.~Alentorn,
\newblock {J. Econ. Dyn. Cont.} {\bf 31}, 2033 (2007)

\bibitem{kyriakopoulos09}
F.~Kyriakopoulos, S.~Thurner, C.~Puhr, and S.W. Schmitz,
\newblock {Eur. Phys. J. B} {\bf 71}, 523 (2009)

\bibitem{ibragimov11}
R.~Ibragimov, D.~Jaffee, and J.~Walden,
\newblock {J. Financ. Econ.} {\bf 99}, 333 (2011)

\bibitem{wagner10}
W.~Wagner,
\newblock {J. Financ. Intermed.} {\bf 19}, 373 (2010)

\bibitem{wagner11}
W.~Wagner,
\newblock {J. Financ.}  {\bf 66}, 1141 (2011)

\bibitem{rashevsky55} N. Rashevsky, \newblock{Bull. Math. Biophys.}, {\bf 17,} 229 (1955)

\bibitem{trucco56} E. Trucco, \newblock{Bull. Math. Biophys.}, {\bf 18}, 129 (1956)

\bibitem{kwapien12}
J. Kwapie$\acute{\mathrm n}$ and S. Dro\.{z}d\.{z}
\newblock {Phys. Rep.} {\bf 515}, 115 (2012)

\bibitem{eisenberg01}
L.~Eisenberg and T.~H. Noe,
\newblock {Manage. Sci.} {\bf 47}, 236 (2001)

\bibitem{brin98} S. Brin, L. Page, \newblock{Comput. Netw.}. {\bf 30}, 107 (1998)

\bibitem{wilhelm07} T. Wilhelm and J. Hollunder, \newblock{Physica A}, {\bf 385}, 385 (2007)

\bibitem{sato10} A.-H. Sato, in \textit{Agent-Based Approaches in Economic and Social Complex Systems VI: Post-Proceedings of The AESCS International Workshop 2009 (Agent-Based Social Systems)}, edited by S.-H. Chen et al. (Springer, Tokyo, 2010), p.3

\bibitem{hurd11}
T.R. Hurd and J.P. Gleeson, \newblock{arXiv: 1110.4312}, (2011)

\bibitem{caccioli12} F. Caccioli, M. Shrestha, C. Moore, J. D. Farmer, \newblock{arXiv:1210.5987} (2012)

\bibitem{soramaki07}
K.~S\"oramaki, M.~Bech, J.~Arnold, R.~Glass, and W.~Beyeler,
\newblock {Physica. A} {\bf 379}, 317 (2007)

\bibitem{iori08}
G.~Iori, G.~De Masi, O.V. Precup, G.~Gabbi, and G.~Caldarelli,
\newblock {J. Econ. Dyn. Cont.} {\bf 32}, 259 (2008)

\bibitem{imakubo10}
K.~Imakubo and Y.~Soejima,
\newblock {Monetary and Economic Studies} {\bf 28}, 107 (2010)



\end{thebibliography}
\end{document}